\documentclass[12pt, a4paper]{article}
\pdfoutput=1
\usepackage{graphicx}
\usepackage{amssymb}
\usepackage{amsmath}
\usepackage{bm}
\usepackage[table,dvipsnames]{xcolor} 
\usepackage{cite}
\usepackage{slashed}
\usepackage{epstopdf}            
\usepackage{epsfig}
\usepackage{here}
\usepackage{comment}
\usepackage{booktabs} 
\usepackage{colortbl} 
\usepackage{wrapfig}
\usepackage{ascmac}
\usepackage{fancybox}  
\usepackage{soul} 
\usepackage{adjustbox}
\usepackage{multirow}
\usepackage{ulem}

\allowdisplaybreaks
\renewcommand{\thefootnote}{\fnsymbol{footnote}}
\numberwithin{equation}{section} 
\def\beq#1\eeq{\begin{align}#1\end{align}}

\renewcommand{\arraystretch}{1.3}
\RequirePackage{xspace}
\usepackage{caption}
\definecolor{BlueViolet}{rgb}{0.2, 0.00, 0.7}
\definecolor{Blue}{rgb}{0.15, 0.00, 0.9}
\definecolor{light_blue}{rgb}{0.15, 0.35, 0.95}
\definecolor{kit_green}{rgb}{0
, 0.58823 
, 0.50980 
}
\usepackage[
colorlinks=true, linkcolor=light_blue,citecolor=light_blue,urlcolor=kit_green]{hyperref} 
\usepackage{xparse}
\NewDocumentCommand{\braket}{mmg}{%
  \IfNoValueTF{#3}
    {\bigl\langle\hspace{0.08em} #1 \hspace{0.08em}\big|\hspace{0.08em} #2 \hspace{0.08em}\bigr\rangle}%
    {\bigl\langle\hspace{0.08em} #1 \hspace{0.08em}\big|\hspace{0.08em} #2 \hspace{0.08em}\big|\hspace{0.08em} #3 \hspace{0.08em}\bigr\rangle}%
}
\newcommand{\ket}[1]{\big|\hspace{0.08em} #1 \hspace{0.08em}\big\rangle}%
\newcommand{\Lb}{\ell}%
\newcommand{\Lc}{\ell^{\hspace{0.04em}\prime}}%
\newcommand{\LbName}{\mathrm{light}\hspace{0.08em}}%
\newcommand{\LcName}{\mathrm{light}^{\prime}\hspace{0.08em}}%

\begin{document}
\sloppy 
\begin{titlepage}
\begin{center}
\hfill{KEK--TH--2754}\\
\vskip .3in

{\Large{\bf Constructing heavy-quark sum rule \\ for $b\to c$ meson and baryon decays}}\\
\vskip .3in

\makeatletter\g@addto@macro\bfseries{\boldmath}\makeatother

{ 
Motoi Endo$^{\rm (a,b,c)}$,
Syuhei Iguro$^{\rm (d,c)}$, 
Satoshi Mishima$^{\rm (e)}$, 
Ryoutaro Watanabe$^{\rm (f)}$
}
\vskip .3in
$^{\rm (a)}${\it KEK Theory Center, IPNS, KEK, Tsukuba 305--0801, Japan}\\\vspace{4pt}
$^{\rm (b)}${\it Graduate Institute for Advanced Studies, SOKENDAI, Tsukuba,\\ Ibaraki 305--0801, Japan} \\\vspace{4pt}
$^{\rm (c)}${\it Kobayashi-Maskawa Institute (KMI) for the Origin of Particles and the Universe, Nagoya University, Nagoya 464--8602, Japan}\\\vspace{4pt}
$^{\rm (d)}${\it Institute for Advanced Research (IAR), Nagoya University,\\ Nagoya 464--8601, Japan}\\\vspace{4pt}
$^{\rm (e)}${\it Department of Liberal Arts, Saitama Medical University, Moroyama,\\ Saitama 350-0495, Japan}\\\vspace{4pt}
$^{\rm (f)}${\it Institute of Particle Physics and Key Laboratory of Quark and Lepton Physics (MOE), Central China Normal University, Wuhan, Hubei 430079, China}
\end{center}
\vskip .15in

\begin{abstract}
We study heavy-hadron semileptonic decays proceeding via $b\to c$ transition, such as $B\to D^{(*)}\tau\bar{\nu}_\tau$ and $\Lambda_b\to\Lambda_c\tau\bar{\nu}_\tau$. 
In the heavy-quark limit, where the heavy-quark symmetry holds, we provide a fundamental framework for heavy-quark sum rules among these decays based on the spin decomposition picture.
The relation holds directly for the squared amplitudes without requiring phase-space integration. 
We then apply this relation to reproduce the sum rule among $B\to D^{(*)}\tau\bar{\nu}_\tau$ and $\Lambda_b\to\Lambda_c\tau\bar{\nu}_\tau$.
Furthermore, we derive new sum rules for $\Omega_b\to \Omega_c^{(*)}$ transitions and those involving excited states, such as $B\to \{D_0^*,D_1^*\}$ and $B\to \{D_1,D_2^*\}$.
\end{abstract}

\end{titlepage}

\setcounter{page}{1}
\renewcommand{\thefootnote}{\#\arabic{footnote}}
\setcounter{footnote}{0}

\hrule
\tableofcontents
\vskip .2in
\hrule
\vskip .4in


\section{Introduction}
\label{sec:intro}

The $b\to c$ semileptonic decays are important probes for testing the Standard Model (SM) and searching for new physics (NP) beyond the SM. 
The anomalies observed in $B\to D^{(*)}\tau\bar{\nu}_\tau$ decays~\cite{HeavyFlavorAveragingSpring2025} have attracted a lot of attention, motivating the study of complementary decay modes with the $b\to c$ transition, such as $\Lambda_b\to\Lambda_c\tau\bar{\nu}_\tau$.

In this context, a sum rule has been proposed among the $B\to D^{(*)}\tau\bar{\nu}_\tau$ and $\Lambda_b\to\Lambda_c\tau\bar{\nu}_\tau$ decay modes~\cite{Blanke:2018yud,Blanke:2019qrx}. 
It connects the ratio observables, $R_{H_c}= \mathrm{BR}(H_b\to H_c\tau\bar{\nu}_\tau)/\mathrm{BR}(H_b\to H_c\hspace{0.08em}l\hspace{0.08em}\bar{\nu}_{l})$ with $l=e,\mu$, through the relation \cite{Endo:2025lvy}
\begin{align}
\frac{R_{\Lambda_c}}{R_{\Lambda_c}^\mathrm{SM}}
\approx
\frac{1}{4}\,
\frac{R_{D}}{R_{D}^\mathrm{SM}} 
+
\frac{3}{4}\,
\frac{R_{D^*}}{R_{D^*}^\mathrm{SM}}
\,,
\label{eq:RSumRule}
\end{align}
which holds even in the presence of NP.
The sum rule provides a tool to test the consistency of the SM predictions and NP effects as well as that of experimental data of the $B\to D^{(*)}\tau\bar{\nu}_\tau$ and $\Lambda_b\to\Lambda_c\tau\bar{\nu}_\tau$ decays. 
The applicability of the sum rule has been investigated in detailed numerical studies in Refs.~\cite{Fedele:2022iib,Duan:2024ayo,Endo:2025fke}.

Besides the sum rule for the $R_{H_c}$ observables, a more fundamental sum rule has been identified in Ref.~\cite{Endo:2025fke}, which connects the differential decay rates of $B\to D^{(*)}\tau\bar{\nu}_\tau$ and $\Lambda_b\to\Lambda_c\tau\bar{\nu}_\tau$, and is given by 
\begin{align}
\frac{\kappa_{\Lambda_c}}{\zeta(w)^2}
=
\frac{2}{1+w} 
\frac{\kappa_{D}+\kappa_{D^*}}{\xi(w)^2} 
\,,
\label{eq:DDRSumRule}
\end{align}
where $\kappa_{H_c}=d\Gamma(H_b\to H_c \tau\bar{\nu}_\tau)/dw$ is the differential decay rate with $w$ being the recoil parameter, and $\xi(w)$ and $\zeta(w)$ denote the leading-order (LO) Isgur--Wise (IW) functions for the transitions of ground-state mesons and those of ground-state baryons, respectively~\cite{Isgur:1989vq,Isgur:1990yhj,Isgur:1990pm,Georgi:1990cx}. 
This sum rule holds exactly in the heavy-quark limit. 
Furthermore, another sum rule was derived for the double-differential decay rates~\cite{Endo:2025cvu}, where $\kappa_{H_c}$ in Eq.~\eqref{eq:DDRSumRule} is replaced by $d^{2}\Gamma(H_b\to H_c \tau\bar{\nu}_\tau)/dw\,d\cos\theta_\tau$ with $\theta_\tau$ being the angle between $H_c$ and $\tau$ in the $H_b$ rest frame. 
The sum rule for the $R_{H_c}$ observables, given in Eq.~\eqref{eq:RSumRule}, can be derived from Eq.~\eqref{eq:DDRSumRule} by taking the zero-recoil limit in addition to the heavy-quark limit~\cite{Endo:2025lvy}.\footnote{ 
In reality, the coefficients in Eq.~\eqref{eq:RSumRule} are slightly modified due to corrections beyond these limits.
}

In this paper, we investigate the theoretical basis of the sum rules by analyzing the underlying structure~\cite{Zalewski:1991xb,Sadzikowski:1993iv} governed by the heavy-quark symmetry~\cite{Isgur:1989vq,Isgur:1990yhj,Nussinov:1986hw,Shifman:1987rj,Eichten:1979pu,Eichten:1980mw,Lepage:1987gg}. 
This analysis is carried out in the heavy-quark limit.
In Ref.~\cite{Endo:2025fke}, the hadronic transitions are evaluated based on the Lorentz covariant formalism of form factors~\cite{Georgi:1990cx,Korner:1987kd,Falk:1990yz,Falk:1991nq}. 
Alternatively, hadron states can be described in the spin decomposition picture, where  wave functions are decomposed into heavy and light states with their spin/angular momenta specified~\cite{Zalewski:1991xb,Sadzikowski:1993iv}.
These two approaches are equivalent in the heavy-quark limit.\footnote{Additionally, in the heavy-quark limit, all bottom (charm) hadron masses are commonly approximated by the bottom (charm) quark mass. 
The mass difference between the bottom and charm quarks is kept nonzero in this paper unless otherwise mentioned explicitly. 
}
Based on the spin decomposition picture, we provide a fundamental framework for sum rules among the squared transition amplitudes of hadron decays by exploiting the orthogonality of the Clebsch--Gordan coefficients.\footnote{In the textbook of Ref.~\cite{Manohar:2000dt} (Secs.~2.3 and 2.4), this spin decomposition picture has been utilized to evaluate a QCD decay process of $D^* \to D \pi$ and probabilities of charm quark hadronizations to $D$ and $D^*$.}  
Since this holds even without performing the phase-space integration, it is more fundamental than those previously identified for (partially) integrated observables such as Eq.~\eqref{eq:DDRSumRule}. 

We then show that the above relation reproduces Eq.~\eqref{eq:DDRSumRule}, namely the sum rule for ground-state transitions among $B\to D^{(*)}$ and $\Lambda_b\to\Lambda_c$.
We also demonstrate that an analogous sum rule holds for other ground-state baryon decays, such as $\Omega_b\to\Omega_c\tau\bar{\nu}_\tau$. 
Furthermore, the relation can be applied to transitions involving excited states. 
We derive a new sum rule for $B\to \{D_0^*,D_1^*\}$ and $B\to \{D_1,D_2^*\}$ transitions.

This paper is organized as follows. 
In Sec.~\ref{sec:GeneralHQsumrule}, we present the derivation of the sum rules.   
We provide an expression for the decay amplitudes in the spin decomposition picture, and then, construct a framework for heavy-quark sum rules for hadron decays.
Explicit examples are given in Sec.~\ref{sec:SumRuleBD} for the $B\to D^{(*)}$ and $\Lambda_b\to\Lambda_c$ transitions, in Sec.~\ref{sec:SumRuleOmega} for the $\Omega_b\to\Omega_c^{(*)}$ transitions, and in Sec.~\ref{sec:SumRuleExcited} for the excited states. 
Our summary and discussion are given in Sec.~\ref{sec:summary}.  
In the Appendices, 
an auxiliary formula is given in Appendix~\ref{sec:auxiliary}, 
explicit formulae for decay amplitudes are listed in Appendix~\ref{sec:decayamplitudes}, 
the definitions of the IW functions are summarized in Appendix~\ref{sec:IWfunctions},
and the Bjorken sum rules are discussed in Appendix~\ref{sec:BjorkenSumRules}.

\section{Construction of heavy-quark sum rule}
\label{sec:GeneralHQsumrule}

Let us consider a singly-heavy hadron $H_Q$, which consists of one heavy quark $Q$ and light degrees of freedom $\Lb$. 
In the heavy-quark limit, the spin-flavor symmetry, known as the heavy-quark symmetry, holds~\cite{Isgur:1989vq,Isgur:1990yhj,Nussinov:1986hw,Shifman:1987rj,Eichten:1979pu,Eichten:1980mw,Lepage:1987gg}. 
The heavy quark behaves as a static color source, analogous to the proton in the hydrogen atom, with the light degrees of freedom playing the role of the electron.
The spin of the heavy quark and the total angular momentum of the light degrees of freedom are separately conserved under QCD interactions within a hadron. 
The heavy-quark symmetry allows hadrons to be described in the spin decomposition picture~\cite{Zalewski:1991xb,Sadzikowski:1993iv}. 
For given hadron spin $\lambda_{H_Q}$ and heavy-quark spin $\lambda_Q$, the hadron state can be decomposed into a product of the heavy-quark spinor $u_Q(\lambda_Q)$ and the state of $\Lb$ as 
\begin{align}
\ket{H_Q(\lambda_{H_Q},\lambda_Q)}
&=
\sum_{\lambda_{\Lb}}\,
\braket
  {s_Q, \lambda_Q\, ; s_{\Lb}, \lambda_{\Lb}}
  {J_{H_Q}, \lambda_{H_Q}}\,
u_Q(\lambda_Q)\,
\ket{\LbName (s_{\Lb}^{P_{\Lb}}, \lambda_{\Lb})}
\,.  
\label{eq:wavefn}
\end{align}
This expression is valid in the heavy-quark limit. 
Here, $J_{H_Q}$, $s_{Q}$ and $s_{\Lb}$ are the total angular momenta of $H_Q$, $Q$ and $\Lb$, respectively,\footnote{$s_{\Lb}$ is the sum of the spin of $\Lb$ and the orbital angular momentum between $Q$ and $\Lb$ in the non-relativistic constituent quark model.} the $\lambda_{H_Q}$, $\lambda_{Q}$ and $\lambda_{\Lb}$ are their projections, $P_{\Lb}$ is the parity of $\Lb$, and $\langle s_Q, \lambda_Q ; s_{\Lb}, \lambda_{\Lb} | J_{H_Q}, \lambda_{H_Q} \rangle$ is the Clebsch--Gordan coefficient. 
The wave functions are normalized as $\bar u_Q(\lambda_Q) u_Q(\lambda_Q) = 2m_{Q}$ and $\braket{\LbName (s_{\Lb}^{P_{\Lb}}, \lambda_{\Lb})}{\LbName (s_{\Lb}^{P_{\Lb}}, \lambda_{\Lb})} = 1$, where $m_Q$ is the heavy-quark mass. 
Under the heavy-quark symmetry, $s_Q=1/2$ is fixed, while $s_{\Lb}$ and $P_{\Lb}$ classify the hadron state. 
Both the heavy-quark spinor and the light-component state implicitly depend on the four-velocity $v$ of the heavy quark, though this dependence is not shown explicitly. 
In the discussion of the $H_b\to H_c$ transition below, we take all $\lambda_i$ to denote projections along the decay axis, defined as the direction of the three-momentum of $H_c$ in the $H_b$ rest frame.

The decay amplitudes for $H_b\to H_c\hspace{0.08em}l\hspace{0.08em}\bar{\nu}_l$ processes with $l=e,\mu,\tau$ are calculated using a low-energy effective Hamiltonian of the form $\mathcal{H}_\mathrm{eff} = \sum_{X} C_X O_X$, where $O_X$ are local operators, and $C_X$ are the corresponding Wilson coefficients. 
We write the local operators as $O_X=(\bar{c}\, \Gamma_X b)(\bar{l}\, \Gamma^{\prime}_X \nu_l)$. 
The decay amplitude for given hadron spins $\lambda_{H_b}$ and $\lambda_{H_c}$ is expressed as \begin{align}
&\mathcal{M}_{H_c}(\lambda_{H_c},\lambda_{H_b})
\nonumber\\
&\hspace{10mm}= 
\sum_{X} C_X
\sum_{\lambda_c,\lambda_b}\,
\braket
  {H_c(\lambda_{H_c},\lambda_c)}
  {\bar{c}\, \Gamma_X b}
  {H_b(\lambda_{H_b},\lambda_b)}\,
\braket
  {l\hspace{0.08em}\bar{\nu}_l}
  {\bar{l}\, \Gamma^{\prime}_X \nu_l}
  {0}\,, 
\label{eq:decayamplitude}
\end{align}
where the velocities of the $H_b$ and $H_c$ hadrons are $v$ and $v'$, respectively. 
The recoil parameter is given by $w = v\cdot v'$.
Using Eq.~\eqref{eq:wavefn}, the matrix element of the hadronic current takes the form, 
\begin{align}
&
\braket
  {H_c(\lambda_{H_c},\lambda_c)}
  {\bar{c}\, \Gamma_X b}
  {H_b(\lambda_{H_b},\lambda_b)}
\nonumber\\[1mm]
&\hspace{5mm}= 
\sum_{\lambda}\,
\braket
  {s_c, \lambda_c\, ; s_{\Lc}, \lambda_{\Lc}}
  {J_{H_c}, \lambda_{H_c}}
\braket
  {s_b, \lambda_b\, ; s_{\Lb}, \lambda_{\Lb}}
  {J_{H_b}, \lambda_{H_b}}
\nonumber\\
&\hspace{17mm}\times 
\bar{u}_c(\lambda_c) \Gamma_X u_b(\lambda_b)\,
\braket
  {\LcName (s_{\Lc}^{P_{\Lc}}, \lambda_{\Lc})}
  {\LbName (s_{\Lb}^{P_{\Lb}}, \lambda_{\Lb})}
\,,  
\label{eq:hadronic}
\end{align}
where 
$\Lc$\,(``$\LcName$\,'') and $\Lb$\,(``$\LbName$'') denote the light degrees of freedom in $H_c$ and $H_b$, respectively.  
In the heavy-quark limit, the projection of the total angular momentum of the light degrees of freedom along the decay axis is conserved~\cite{Politzer:1990ps}, and hence, 
we denote $\lambda_{\Lc} = \lambda_{\Lb} \equiv \lambda$.
The matrix element of the light degrees of freedom, which corresponds to the overlap of the light-component wave functions, satisfies the following relation~\cite{Politzer:1990ps,Zalewski:1991xb}:
\begin{align}
&
\braket
  {\LcName (s_{\Lc}^{P_{\Lc}}, -\lambda)}
  {\LbName (s_{\Lb}^{P_{\Lb}}, -\lambda)}
\nonumber\\
&\hspace{5mm}
=
P_{\Lb}
P_{\Lc}
(-1)^{s_{\Lb}-s_{\Lc}}
\braket
  {\LcName (s_{\Lc}^{P_{\Lc}}, \lambda)}
  {\LbName (s_{\Lb}^{P_{\Lb}}, \lambda)}
\,. 
\label{eq:Lmatrix}
\end{align}
Since $\lambda$ is a common projection of $s_{\Lc}$ and $s_{\Lb}$, both $s_{\Lc}$ and $s_{\Lb}$ must be either integers or half-integers.
Combining Eqs.~\eqref{eq:decayamplitude} and \eqref{eq:hadronic}, the decay amplitude is expressed as
\begin{align}
\mathcal{M}_{H_c}(\lambda_{H_c},\lambda_{H_b})
&= 
\sum_{\lambda_c,\lambda_b}
\sum_{\lambda}\,
\braket
  {s_c, \lambda_c\, ; s_{\Lc}, \lambda}
  {J_{H_c}, \lambda_{H_c}}
\braket
  {s_b, \lambda_b\, ; s_{\Lb}, \lambda}
  {J_{H_b}, \lambda_{H_b}}
\nonumber \\
&\hspace{5mm} \times 
M_{\lambda_c,\lambda_b}\,
\braket
  {\LcName (s_{\Lc}^{P_{\Lc}}, \lambda)}
  {\LbName (s_{\Lb}^{P_{\Lb}}, \lambda)}
\,,
\label{eq:decayamplitude2}
\end{align}
where $M_{\lambda_c,\lambda_b}$ is the matrix element for the quark-level transition, including the leptonic matrix element, 
\begin{align}
M_{\lambda_c,\lambda_b}
=
\sum_{X} C_X\,
\bar{u}_c(\lambda_c) \Gamma_X u_b(\lambda_b)\,
\braket
  {l\hspace{0.08em}\bar{\nu}_l}
  {\bar{l}\, \Gamma^{\prime}_X \nu_l} 
  {0}
\,.
\label{eq:quarklevel}
\end{align}
Hereafter, we abbreviate $M_{1/2,1/2}$ as $M_{++}$, $M_{1/2,-1/2}$ as $M_{+-}$, and so on.
We stress that $M_{\lambda_c,\lambda_b}$ is independent of the hadron species $H_b$ and $H_c$, which is a key feature to construct the sum rule, while $\langle \LcName (s_{\Lc}^{P_{\Lc}}, \lambda) | \LbName (s_{\Lb}^{P_{\Lb}}, \lambda)\rangle$ corresponds to $H_b \to H_c$ transition form factors, as we will see later explicitly. 
Both matrix elements of the heavy part, $M_{\lambda_c,\lambda_b}$, and that of the light components, $\langle \LcName (s_{\Lc}^{P_{\Lc}}, \lambda) | \LbName (s_{\Lb}^{P_{\Lb}}, \lambda)\rangle$, depend on the recoil parameter $w$ implicitly.

By squaring the amplitude in Eq.~\eqref{eq:decayamplitude2} and summing over $\lambda_{H_b}$ and $\lambda_{H_c}$, together with the sum over possible $J_{H_c}$, we obtain 
\begin{align}
&
\sum_{J_{H_c}}
\frac{1}{2J_{H_b}+1}
\,\sum_{\lambda_{H_c},\lambda_{H_b}}\,\!\!\!
\big|\mathcal{M}_{H_c}(\lambda_{H_c},\lambda_{H_b})\big|^2
\nonumber\\
&\hspace{5mm}= 
\frac{1}{2}
\sum_{\lambda_c,\lambda_b} 
\big| M_{\lambda_c,\lambda_b} \big|^2\,
\frac{2}{2J_{H_b}+1}
\sum_{\lambda} 
\big|
\braket
  {\tfrac12, \lambda_b\, ; s_{\Lb}, \lambda}
  {J_{H_b}, \lambda_{b}+\lambda}
\big|^2\,
\nonumber\\
&\hspace{10mm}\times
\big|
\braket
  {\LcName (s_{\Lc}^{P_{\Lc}}, |\lambda|)}
  {\LbName (s_{\Lb}^{P_{\Lb}}, |\lambda|)}
\big|^2
\,,
\label{eq:amplitudeSquared}
\end{align}
where the factor $1/(2J_{H_b}+1)$ is introduced to make an average over the initial-hadron spin. 
Here, we have used the selection rule $\langle j_1,m_1 ; j_2,m_2 | J, M \rangle \propto \delta_{m_1+m_2,M}$ and the orthogonality relation,
\begin{align}
\sum_{J_{H_c}}
\sum_{\lambda_{H_c}}
\braket
  {\tfrac12, \lambda_c\, ; s_{\Lc}, \lambda}
  {J_{H_c}, \lambda_{H_c}}
\braket
  {J_{H_c}, \lambda_{H_c}}
  {\tfrac12, \lambda'_c\, ; s_{\Lc}, \lambda'}
=
\delta_{\lambda_c,\lambda'_c}
\delta_{\lambda,\lambda'}
\,.
\label{eq:CG-orthogonality}
\end{align}
It is noticed that the summation over the final-state spin $J_{H_c}$ allows us to use the orthogonality of the Clebsch--Gordan coefficients.  
If instead the summation is taken over the initial-state spin $J_{H_b}$, the presence of the factor $1/(2J_{H_b}+1)$ prevents the use of the orthogonality relation. 

Let us describe the squared hadron decay amplitude normalized with the squared form factor as 
\begin{align}
\Delta(H_b\to H_c\hspace{0.08em}l\hspace{0.08em}\bar{\nu}_l) \equiv
\frac{
\displaystyle
\sum_{J_{H_c}}
\frac{1}{2J_{H_b}+1}
\sum_{\lambda_{H_c},\lambda_{H_b}}\!\!\!
\big|\mathcal{M}_{H_c}(\lambda_{H_c},\lambda_{H_b})\big|^2
}{\phantom{\Bigg|}
\displaystyle
\frac{1}{2s_{\Lb}+1}
\sum_{\lambda} 
\big|
\braket
  {\LcName (s_{\Lc}^{P_{\Lc}}, |\lambda|)}
  {\LbName (s_{\Lb}^{P_{\Lb}}, |\lambda|)}
\big|^2
}
\,,
\label{eq:Delta}
\end{align}
where the summation over $J_{H_c}$ corresponds to that over all possible hadrons $H_c$ which can be formed for given $s_{\Lc}^{P_{\Lc}}$. 
From Eq.~\eqref{eq:amplitudeSquared} by using an auxiliary formula in  Eq.~\eqref{eq:CGrelation}, we arrive at the relation, 
\begin{align}
\Delta(H_b\to H_c\hspace{0.08em}l\hspace{0.08em}\bar{\nu}_l) 
=
\frac{1}{2}
\sum_{\lambda_c,\lambda_b} 
\big| M_{\lambda_c,\lambda_b} \big|^2\,
\,. 
\label{eq:AmpSquaredGeneral}
\end{align}
The right-hand side corresponds to the squared amplitude of the partonic $b\to c\hspace{0.08em}l\hspace{0.08em}\bar{\nu}_l$ decay averaged over the initial-quark spin.
Hence, based on the spin decomposition picture, the hadron decay amplitude, which is given by the numerator of $\Delta(H_b\to H_c\hspace{0.08em}l\hspace{0.08em}\bar{\nu}_l)$, is expressed by the partonic decay amplitude multiplied by a form factor in the heavy-quark limit.

Since $M_{\lambda_c,\lambda_b}$ is independent of the hadron species $H_b$ and $H_c$ as stressed above, Eq.~\eqref{eq:AmpSquaredGeneral} leads to a framework for sum rules among various heavy-hadron decay channels,\footnote{In Ref.~\cite{Amano:2021spn}, a sum rule was derived for the squared decay amplitudes of $\bar{B}_{s}^{0}\to D_{s}^{(*)+}l\hspace{0.08em}\bar{\nu}_l$ and $\Lambda_{b}\to\Lambda_{c}\hspace{0.08em}l\hspace{0.08em}\bar{\nu}_l$, under the conjecture of a dynamical supersymmetry between the $\bar{s}$ constituent quark and the $ud$ scalar diquark.}
\begin{align}
\Delta(H_b\to H_c\hspace{0.08em}l\hspace{0.08em}\bar{\nu}_l) 
=
\Delta(H_b'\to H_c'\hspace{0.08em}l\hspace{0.08em}\bar{\nu}_l) 
=
\cdots
=
\frac{1}{2}
\sum_{\lambda_c,\lambda_b} 
\big| M_{\lambda_c,\lambda_b} \big|^2\,
\,.
\label{eq:GeneralSumRule}
\end{align}
This relation holds directly for the squared amplitudes even without requiring phase-space integration. 
From this relation, we can construct sum rules for decay rates as demonstrated in the next section. 
We note that the sum rules hold separately for each spin amplitude, namely for the terms $|M_{++}|^2$, $|M_{--}|^2$, $|M_{-+}|^2$ and $|M_{+-}|^2$. 
Furthermore, the sum rule holds irrespective of the presence of NP contributions, which enter through $M_{\lambda_c,\lambda_b}$. 

In addition, the sum rule derived above can be extended beyond semileptonic transitions. 
In fact, by replacing the lepton current in Eq.~\eqref{eq:decayamplitude2} with a light-quark current, one can derive a sum rule for hadronic decays. 
However, this requires that the matrix element of the light-quark current is factorized from that of the light component inside the heavy hadron. 
This condition is approximately satisfied in Class-I $B$-meson decays into heavy-light final states, such as $\bar{B}^0 \to D^{(*)+} K^-$ and $\bar{B}_s^0 \to D_s^{(*)+} \pi^-$.

\section{Explicit examples}
\label{sec:SumRule}

\subsection{Sum rule for $B\to D^{(*)}$ and $\Lambda_b\to\Lambda_c$}
\label{sec:SumRuleBD}

Let us consider the $B\to D^{(*)}$ and $\Lambda_b\to\Lambda_c$ transitions, where $J_{H_b}=0$, $J_{H_c}=\{0,1\}$ and $s_{\Lb}^{P_{\Lb}}=s_{\Lc}^{P_{\Lc}}=\tfrac12^{-}$ for the former, and $J_{H_b}=J_{H_c}=\tfrac12$ and $s_{\Lb}^{P_{\Lb}}=s_{\Lc}^{P_{\Lc}}=0^{+}$ for the latter.  
The quantum numbers of the singly-heavy ground-state hadrons are listed in Table~\ref{tab:hadrons}.

\begin{table}[t]
  \renewcommand{\arraystretch}{1.3}
  \centering
  \begin{tabular}{c|cc|c|c}
    \hline
    Content 
    & $I\,(J^{P})$ & $s_{\Lb}^{P_{\Lb}}$ 
    & 
    $b$ hadron & 
    $c$ hadron 
    \\
    \hline
    \multirow{2}{*}{$Q\bar q$}
    & $\tfrac12\,(0^{-})$
    & \multirow{2}{*}{$\tfrac12^{-}$} 
    & $B^-$, $\bar{B}^0$
    & $D^{0}$, $D^{+}$
    \\
    & $\tfrac12\,(1^{-})$
    & 
    & $B^{*-}$, $\bar{B}^{*0}$
    & $D^{*0}$, $D^{*+}$
    \\
    \hline
    \multirow{2}{*}{$Q\bar s$}
    & $0\,(0^{-})$
    & \multirow{2}{*}{$\tfrac12^{-}$} 
    & $\bar{B}_{s}^{0}$
    & $D_{s}^{+}$
    \\
    & $0\,(1^{-})$
    & 
    & $\bar{B}_{s}^{*0}$
    & $D_{s}^{*+}$
    \\
    \hline
    $Q[ud]$
    & $0\,(\tfrac12^{+})$ & $0^{+}$ 
    & $\Lambda_b^{0}$ 
    & $\Lambda_c^{+}$ 
    \\
    \hline
    $Q[qs]$ 
    & $\tfrac12\,(\tfrac12^{+})$ & $0^{+}$ 
    & $\Xi_b^{0}$, $\Xi_b^{-}$
    & $\Xi_c^{+}$, $\Xi_c^{0}$
    \\
    \hline
    \multirow{2}{*}{$Q\{qs\}$}
    & $\tfrac12\,(\tfrac12^{+})$ & \multirow{2}{*}{$1^{+}$}  
    & $\Xi_b^{\prime 0}$, $\Xi_b^{\prime -}$
    & $\Xi_c^{\prime +}$, $\Xi_c^{\prime 0}$
    \\
    & $\tfrac12\,(\tfrac32^{+})$ & 
    & $\Xi_b^{*0}$, $\Xi_b^{*-}$
    & $\Xi_c^{*+}$, $\Xi_c^{*0}$
    \\
    \hline
    \multirow{2}{*}{$Q\{qq\}$}
    & $1\,(\tfrac12^{+})$ & \multirow{2}{*}{$1^{+}$}  
    & $\Sigma_b^{+}$, $\Sigma_b^{0}$, $\Sigma_b^{-}$
    & $\Sigma_c^{++}$, $\Sigma_c^{+}$, $\Sigma_c^{0}$
    \\
    & $1\,(\tfrac32^{+})$ & 
    & $\Sigma_b^{*+}$, $\Sigma_b^{*0}$, $\Sigma_b^{*-}$
    & $\Sigma_c^{*++}$, $\Sigma_c^{*+}$, $\Sigma_c^{*0}$
    \\
    \hline
    \multirow{2}{*}{$Q\{ss\}$}
    & $0\,(\tfrac12^{+})$ & \multirow{2}{*}{$1^{+}$}  
    & $\Omega_b^{-}$
    & $\Omega_c^{0}$
    \\
    & $0\,(\tfrac32^{+})$ & 
    & $\Omega_b^{*-}$
    & $\Omega_c^{*0}$
    \\
    \hline
  \end{tabular}
  \caption{Ground states of bottom ($Q=b$) and charm ($Q=c$) hadrons, where $q$ denotes $u$ or $d$. 
  The ``Content'' column refers to the quark content in the constituent quark model, where the brackets $[\,]$ indicate a ``good'' diquark, while the braces $\{\,\}$ indicate a ``bad'' diquark. 
  $I$ and $J^P$ are the isospin and the spin-parity of the hadron, respectively, and $s_{\Lb}^{P_{\Lb}}$ is the spin-parity of the light component.  Quarks (antiquarks) are assigned positive (negative) parity by convention.
  }
  \label{tab:hadrons}
\end{table}

Using Eq.~\eqref{eq:GeneralSumRule}, we obtain the sum rule 
\begin{align}
\Delta(B\to D^{(*)}\hspace{0.08em}l\hspace{0.08em}\bar{\nu}_l) 
=
\Delta(\Lambda_b\to \Lambda_c\hspace{0.08em}l\hspace{0.08em}\bar{\nu}_l) 
=
\frac{1}{2}
\sum_{\lambda_c,\lambda_b} 
\big| M_{\lambda_c,\lambda_b} \big|^2\,
\,,
\end{align}
or it is explicitly shown as
\begin{align}
\frac{
\displaystyle
\big|
  \mathcal{M}_{D}
\big|^2
+
\sum_{\lambda_{D^*}}
\big|
  \mathcal{M}_{D^*}(\lambda_{D^*})
\big|^2
}{
\braket
  {\bar{q} (\tfrac{1}{2}^{-}, \tfrac{1}{2})}
  {\bar{q} (\tfrac{1}{2}^{-}, \tfrac{1}{2})}^2
}
&=
\frac{
\displaystyle
\frac12
\sum_{\lambda_{\Lambda_{c}},\lambda_{\Lambda_{b}}}
\big|
  \mathcal{M}_{\Lambda_c}(\lambda_{\Lambda_{c}},\lambda_{\Lambda_{b}})
\big|^2
}{
\braket
  {qq (0^{+},0)}
  {qq (0^{+},0)}^2
}
\,. 
\label{eq:sumrule}
\end{align}
As in the constituent quark model, the light component on the left-hand side is effectively described by an antiquark $\bar{q} = \bar{u}$ or $\bar{d}$, while that on the right-hand side is by a spinless and parity-even diquark, the so-called ``good'' diquark. 
The full expressions of the decay amplitudes are given in Appendix~\ref{sec:decayamplitudes}. 
Since the only difference between the $B\to D^{(*)}$ and $\Lambda_b\to\Lambda_c$ transitions lies in the structure of the light components, the equality in Eq.~\eqref{eq:sumrule} reflects that the underlying heavy-quark transitions are identical.
Furthermore, due to the orthogonality of the Clebsch--Gordan coefficients, the contributions from the $B\to D$ and $B\to D^*$ modes must be summed on the left-hand side, reflecting the fact that $D$ and $D^*$ form a heavy-quark spin doublet.

The LO mesonic IW function $\xi(w)$ is expressed in terms of the matrix element of the light component as~\cite{Sadzikowski:1993iv}
\begin{align}
\xi(w)
&=
\sqrt{\frac{2}{w+1}}\,
\braket
  {\bar{q} (\tfrac{1}{2}^{-}, \tfrac{1}{2})}
  {\bar{q} (\tfrac{1}{2}^{-}, \tfrac{1}{2})}
\,.
\label{eq:xi}
\end{align}
Similarly, the baryonic IW function $\zeta(w)$ is written as
\begin{align}
\zeta(w)
&=
\braket
  {qq (0^{+},0)}
  {qq (0^{+},0)}
\label{eq:zeta}
\,. 
\end{align}
More details of these relations are given in Appendix~\ref{sec:IWfunctions}. 
Using Eqs.~\eqref{eq:xi} and \eqref{eq:zeta}, the sum rule in Eq.~\eqref{eq:sumrule} can be rewritten as 
\begin{align}
\frac{2}{1+w}
\frac{
\displaystyle
\big|
  \mathcal{M}_{D}
\big|^2
+
\sum_{\lambda_{D^*}}
\big|
  \mathcal{M}_{D^*}(\lambda_{D^*})
\big|^2
\vphantom{\Big|}
}{
\xi(w)^2
}
&=
\frac{
\displaystyle
\frac12
\sum_{\lambda_{\Lambda_{c}},\lambda_{\Lambda_{b}}}
\big|
  \mathcal{M}_{\Lambda_c}(\lambda_{\Lambda_{c}},\lambda_{\Lambda_{b}})
\big|^2
\vphantom{\bigg|}
}{
\zeta(w)^2
}
\,. 
\label{eq:sumruleBLambda}
\end{align}
From this relation, Eq.~\eqref{eq:DDRSumRule} is reproduced by multiplying a kinematic factor necessary to obtain the decay rates from the squared amplitudes appearing in the numerators of Eq.~\eqref{eq:Delta}.
Additionally, the phase-space integration is performed over the angular distributions of the final-state particles, while $w$ is kept unintegrated.
It is noticed that, in the heavy-quark limit, since the hadron masses are commonly approximated by the corresponding heavy-quark masses, these factors are universal in all transitions. 
Also, the form factors are functions of $w$ but independent of the angular distributions.
We stress that the sum rule derived here is more fundamental than Eq.~\eqref{eq:DDRSumRule}, because the latter is obtained by performing an integration over the phase space.

\subsection{Sum rule for $\Omega_b\to\Omega_c^{(*)}$}
\label{sec:SumRuleOmega}

In the previous subsection, we discussed the decay of the ground-state bottom baryon $\Lambda_{b}$ to the ground-state charm baryon $\Lambda_{c}$, both characterized by $J^P = \tfrac12^{+}$. 
In this case, the light component in the baryons is effectively described by a spinless ``good'' diquark. 
We now turn to the case where the light component is described by a spin-one and parity-even diquark, the so-called ``bad'' diquark. 
Here, we derive a sum rule for the $\Omega_{b}\to\Omega_{c}^{(*)}$ transitions, where $\Omega_{b}$ and $\Omega_{c}$ have $J^P = \tfrac12^{+}$, while $\Omega_{c}^{*}$ has $J^P = \tfrac32^{+}$. 
The derived formula can be applied to other transitions involving the ``bad'' diquark, {\it i.e.}, $\Xi_{b}^{\prime}\to\Xi_{c}^{\prime}(\Xi_{c}^{*})$ and $\Sigma_{b}\to\Sigma_{c}^{(*)}$.

In the case of the $\Omega_{b}\to \Omega_{c}^{(*)}$ transitions, there are two LO IW functions, $\xi_1(w)$ and $\xi_2(w)$~\cite{Isgur:1990pm,Georgi:1990cx}. 
We find that they are related to the matrix elements of the spin-one diquark as 
\begin{align}
\xi_1(w) 
&=
\braket
  {ss(1^{+},1)}
  {ss(1^{+},1)}
\,,
\label{eq:IW_xi1}
\\
\xi_2(w) 
&=
\frac{1}{w^2-1}
\Big(
w\,
\braket
  {ss(1^{+},1)}
  {ss(1^{+},1)}
-
\braket
  {ss(1^{+},0)}
  {ss(1^{+},0)}
\Big)
\,,
\label{eq:IW_xi2}
\end{align}
where $\xi_1(w)$ is normalized as $\xi_1(1)=1$, and the derivation of these relations is given in Appendix~\ref{sec:IWfunctions}. 
The denominator in Eq.~\eqref{eq:Delta} is expressed as\footnote{The combination of the IW functions in Eq.~\eqref{eq:relationOmega} is the same as that appearing in the ground-state contribution to the Bjorken sum rule for the semileptonic $\Omega_{b}$ decays~\cite{Xu:1993mj}. } 
\begin{align}
&\frac13\,
\Big(
\braket
  {ss(1^{+},0)}
  {ss(1^{+},0)}^2
+
2\,
\braket
  {ss(1^{+},1)}
  {ss(1^{+},1)}^2
\Big)
\nonumber\\
&\hspace{10mm}
=
\frac{2+w^2}{3}\,
\xi_1(w)^2
+
\frac{(w^2-1)^2}{3}\,
\xi_2(w)^2
-
\frac{2w(w^2-1)}{3}\,
\xi_1(w)\,
\xi_2(w)
\,.
\label{eq:relationOmega}
\end{align}
If the equality $\langle ss(1^{+},1) | ss(1^{+},1) \rangle = \langle ss(1^{+},0) | ss(1^{+},0) \rangle$ holds, we have the relation $\xi_2(w) = \xi_1(w)/(w+1)$, which was derived in the spectator-quark model~\cite{Korner:1994nh}, in the large $N_c$ limit~\cite{Chow:1994ni}, and in the relativistic quark model~\cite{Ebert:2006rp}. 

From Eq.~\eqref{eq:GeneralSumRule}, we find the sum rule for the $\Omega_{b}\to \Omega_{c}^{(*)}$ transitions $\Delta(\Omega_{b}\to \Omega_{c}^{(*)}\hspace{0.08em}l\hspace{0.08em}\bar{\nu}_l)$ as
\begin{align}
\Delta(\Omega_{b}\to \Omega_{c}^{(*)}\hspace{0.08em}l\hspace{0.08em}\bar{\nu}_l)
&=
\frac{
\displaystyle
\sum_{H_c=\Omega_{c},\Omega_{c}^{*}}
\frac12
\sum_{\lambda_{H_c},\lambda_{\Omega_{b}}}
\big|
  \mathcal{M}_{H_c}(\lambda_{H_c},\lambda_{\Omega_{b}})
\big|^2
\vphantom{\bigg|}
}{
\vphantom{\Bigg|}
\displaystyle
\frac{2+w^2}{3}\,
\xi_1(w)^2
+
\frac{(w^2-1)^2}{3}\,
\xi_2(w)^2
-
\frac{2w(w^2-1)}{3}\,
\xi_1(w)\,
\xi_2(w)
}
\,,
\nonumber\\
&=
\frac{1}{2}
\sum_{\lambda_c,\lambda_b} 
\big| M_{\lambda_c,\lambda_b} \big|^2.
\end{align}

\subsection{Sum rule for excited states}
\label{sec:SumRuleExcited}

\begin{table}[tp]
  \renewcommand{\arraystretch}{1.3}
  \centering
  \begin{tabular}{c|cc|c}
    \hline
    Content 
    & $I\,(J^{P})$ & $s_{\Lb}^{P_{\Lb}}$ 
    & 
    $c$ hadron 
    \\
    \hline
    \multirow{4}{*}{$c\bar q$}
    & $\tfrac12\,(0^{+})$
    & \multirow{2}{*}{$\tfrac12^{+}$} 
    & $D_{0}^{*0}$, $D_{0}^{*+}$
    \\
    & $\tfrac12\,(1^{+})$
    & 
    & $D_{1}^{*0}$,$D_{1}^{*+}$
    \\
    \cline{2-4}
    & $\tfrac12\,(1^{+})$
    & \multirow{2}{*}{$\tfrac32^{+}$} 
    & $D_{1}^{0}$, $D_{1}^{+}$
    \\
    & $\tfrac12\,(2^{+})$
    & 
    & $D_{2}^{*0}$, $D_{2}^{*+}$
    \\
    \hline  \end{tabular}
  \caption{Four lightest excited states of charm mesons.}
  \label{tab:charmExcited}
\end{table}
Equation~\eqref{eq:GeneralSumRule} also provides sum rules for transitions involving excited states. 
For example, the $B\to D_0^*$ and $B\to D_1^*$ transitions correspond to $J_{H_b}=0$, $s_{\Lb}^{P_{\Lb}}=\tfrac12^{-}$ and $s_{\Lc}^{P_{\Lc}}=\tfrac12^{+}$, while the $B\to D_1$ and $B\to D_2^*$ correspond to $J_{H_b}=0$, $s_{\Lb}^{P_{\Lb}}=\tfrac12^{-}$ and $s_{\Lc}^{P_{\Lc}}=\tfrac32^{+}$, where the quantum numbers of the excited charm mesons are summarized in Table~\ref{tab:charmExcited}. 
The IW functions for these transitions, $\tau_{1/2}(w)$ and $\tau_{3/2}(w)$~\cite{Isgur:1990jf}, are given by~\cite{Veseli:1995fr}\footnote{
The matrix element $\braket{\LcName (s_{\Lc}^{P_{\Lc}}, \lambda)}{\LbName (s_{\Lb}^{P_{\Lb}}, \lambda)}$ vanishes for $s_{\Lc}^{P_{\Lc}} \neq s_{\Lb}^{P_{\Lb}}$ at zero recoil $w \to 1$, because $H_b$ and $H_c$ belong to different multiplets of the heavy-quark spin symmetry. } 
\begin{align}
\tau_{1/2}(w)
&=
\frac{1}{\sqrt{2(w-1)}}\,
\braket
  {\bar{q} (\tfrac{1}{2}^{+}, \tfrac{1}{2})}
  {\bar{q} (\tfrac{1}{2}^{-}, \tfrac{1}{2})}
\,,
\label{eq:IW_tau12}
\\
\tau_{3/2}(w)
&=
-
\frac{1}{(w+1)}
\frac{1}{\sqrt{w-1}}\,
\braket
  {\bar{q} (\tfrac{3}{2}^{+}, \tfrac{1}{2})}
  {\bar{q} (\tfrac{1}{2}^{-}, \tfrac{1}{2})}
\,. 
\label{eq:IW_tau32}
\end{align}
As in the case of the ground-state transitions, we obtain the sum rule for $\Delta(B\to \{D_0^*,D_1^*\}\hspace{0.08em}l\hspace{0.08em}\bar{\nu}_l)$ and $\Delta(B\to \{D_1,D_2^*\}\hspace{0.08em}l\hspace{0.08em}\bar{\nu}_l)$ as 
\begin{align}
\Delta(B\to \{D_0^*,D_1^*\}\hspace{0.08em}l\hspace{0.08em}\bar{\nu}_l)
=
\Delta(B\to \{D_1,D_2^*\}\hspace{0.08em}l\hspace{0.08em}\bar{\nu}_l)
=
\frac{1}{2}
\sum_{\lambda_c,\lambda_b} 
\big| M_{\lambda_c,\lambda_b} \big|^2,
\end{align}
where 
\begin{align}
\Delta(B\to \{D_0^*,D_1^*\}\hspace{0.08em}l\hspace{0.08em}\bar{\nu}_l)
&=
\frac{\displaystyle
\sum_{H_c=D_0^*,D_1^*}
\sum_{\lambda_{H_c}}
\big|
  \mathcal{M}_{H_c}(\lambda_{H_c})
\big|^2
\vphantom{\bigg|}
}{\displaystyle
2(w - 1)\,
\tau_{1/2}(w)^2
}
\,,
\\
\Delta(B\to \{D_1,D_2^*\}\hspace{0.08em}l\hspace{0.08em}\bar{\nu}_l)
&=
\frac{\displaystyle
\sum_{H_c=D_1,D_2^*}
\sum_{\lambda_{H_c}}
\big|
  \mathcal{M}_{H_c}(\lambda_{H_c})
\big|^2
\vphantom{\bigg|}
}{\displaystyle
(w + 1)^2(w - 1)\,
\tau_{3/2}(w)^2
}
\,.
\end{align}

\section{Summary and discussion}
\label{sec:summary}

In this paper, we clarified the theoretical structure underlying the sum rules for the $b\to c$ semileptonic hadron decays such as those among $B\to D^{(*)}\tau\bar{\nu}_\tau$ and $\Lambda_b\to\Lambda_c\tau\bar{\nu}_\tau$. 
The analysis was carried out in the heavy-quark limit, where the heavy-quark symmetry holds.
In this limit, the spin/angular-momentum structure of the decay amplitudes becomes transparent by using the spin decomposition picture.
Adopting this expression, we constructed a fundamental framework for heavy-quark sum rules for heavy-hadron decays by exploiting the orthogonality of the Clebsch--Gordan coefficients.

The sum rules arise from the fact that the differences among the transitions are encoded solely in the matrix element of the light degrees of freedom, namely in the form factors, whereas the quark-level structure of the heavy-quark transition is universal. 
Therefore, by normalizing the squared amplitudes with the form factors, they become independent of the hadron species involved in the decay, thereby yielding the sum rules. 
Moreover, the relation holds directly among the squared amplitudes, {\it i.e.}, without performing the phase-space integration. 

We then applied the above relation to reproduce those for $B\to D^{(*)}\tau\bar{\nu}_\tau$ and $\Lambda_b\to\Lambda_c\tau\bar{\nu}_\tau$.
From the spin structure of the sum rule, we clarified that the $B\to D$ contribution alone is not sufficient, and the $B\to D^{*}$ channel must also be included. 
Furthermore, we extended the analysis to the $\Omega_b$ transitions and derived a new sum rule relating $\Omega_b\to\Omega_c^{(*)}$ to $B\to D^{(*)}$ and $\Lambda_b\to\Lambda_c$.
Analogous to the $B\to D$ and $B\to D^{*}$ case, both $\Omega_b\to\Omega_c$ and $\Omega_b\to\Omega_c^*$ channels are included in the sum rule. 
Although precise experimental data are not yet available, it would be of interest to investigate violations of the sum rule beyond the heavy-quark limit and to predict the $\Omega_b\to\Omega_c^{(*)}$ transitions from other channels using the sum rule. 
As the third example, we also constructed a sum rule for transitions involving excited states, $B\to \{D_0^*,D_1^*\}$ and $B\to \{D_1,D_2^*\}$.

It should be emphasized that the equality of the sum rules holds only in the heavy-quark limit, and deviations arise because the limit is broken for realistic hadrons.
Although the corrections have been evaluated for the sum rule among $B\to D^{(*)}$ and $\Lambda_b\to\Lambda_c$~\cite{Endo:2025lvy, Endo:2025fke}, they remain unexplored for others. 
Their detailed analyses are beyond the scope of this paper and will be performed elsewhere.

There are a variety of further applications. 
For instance, assuming the flavor $SU(3)$ symmetry, one can construct analogous sum rules involving $B_s\to D_s^{(*)}$ and so on.
Naively, the $SU(3)$ symmetry breaking amounts to $\mathcal{O}(10\%)$ corrections~\cite{Bordone:2019guc}, and hence, could not be negligible.
However, a detailed analysis is necessary to evaluate its effects on the sum rules.
Moreover, the sum rules can be extended beyond semileptonic transitions, {\it e.g.}, to Class-I hadronic two-body decays such as $\bar{B}^0 \to D^{(*)+} K^-$ and $\bar{B}_s^0 \to D_s^{(*)+} \pi^-$. 
A detailed study for these cases is also left for future work. 

On the experimental side, precise data are not yet available for many decay channels. 
The current and future experiments such as Belle II, LHCb, and Tera-Z are expected to measure them with high precisions (see, e.g., Refs.~\cite{Belle-II:2018jsg, Belle-II:2022cgf, ATLAS:2025lrr, Bernlochner:2021vlv, Ho:2022ipo, Ai:2024nmn}). 
For instance, LHCb is projected to determine $R_{D_s^{(*)}}$ and $R_{D_1}$ at percent levels~\cite{Bernlochner:2021vlv}, while the precision for $R_{D_s^{(*)}}$ at Tera-Z could reach $\mathcal{O}(0.1)\%$, although systematic uncertainties have not been taken into account~\cite{Ho:2022ipo, Ai:2024nmn}.
These experimental results would be applied to the sum rules in future.
Since many decay channels remain unexplored, further studies are required. 

Finally, as an application of the spin decomposition picture of the decay amplitudes, we reproduced the Bjorken sum rules based on this picture in Appendix~\ref{sec:BjorkenSumRules}.

\section*{Acknowledgements}
This work is supported by JSPS KAKENHI Grant Numbers 22K21347 [M.E. and S.I.], 24K07025 [S.M.], 24K22879 [S.I.], 24K23939 [S.I.] and 25K17385 [S.I.]. 
The work is also supported by JPJSCCA20200002 and the Toyoaki scholarship foundation [S.I.].
We also appreciate KEK-KMI joint appointment program [M.E. and S.I.], which accelerated this project. 
\appendix

\section{Auxiliary relation}
\label{sec:auxiliary}

In this Appendix, we derive the auxiliary relation that is used to obtain Eq.~\eqref{eq:AmpSquaredGeneral}.

The square of the Clebsch--Gordan coefficients appearing in Eq.~\eqref{eq:amplitudeSquared} takes the following form: 
\begin{align}
\big|
\braket
  {\tfrac12, \lambda_b\, ; s_{\Lb}, \lambda}
  {J_{H_b}, \lambda_b+\lambda}
\big|^2
=
\left\{
\begin{array}{ll}
\displaystyle
\frac{s_{\Lb} \pm \lambda+1}{2s_{\Lb}+1}
&\quad\mathrm{for}\ J_{H_b} = s_{\Lb} + \tfrac12\,,\
\lambda_b=\pm\tfrac12\,,
\\[4mm]
\displaystyle
\frac{s_{\Lb}\mp\lambda}{2s_{\Lb}+1}
&\quad\mathrm{for}\ J_{H_b} = s_{\Lb} - \tfrac12\,,\
\lambda_b=\pm\tfrac12\,. 
\end{array}
\right.
\end{align}
In particular, for $s_{\Lb}=0$, this expression reduces to
\begin{align}
\big|
\braket
  {\tfrac12, \lambda_b\, ; s_{\Lb}, \lambda}
  {J_{H_b}, \lambda_b+\lambda}
\big|^2
=
\big|
\braket
  {\tfrac12, \lambda_b\, ; 0, 0}
  {\tfrac12, \lambda_b}
\big|^2
=
1
\,. 
\end{align}
Using these expressions, we obtain 
\begin{align}
&
\frac{2}{2J_{H_b}+1}
\sum_{\lambda=-J}^{J}
\big|
\braket
  {\tfrac12, \lambda_b\, ; s_{\Lb}, \lambda}
  {J_{H_b}, \lambda_b+\lambda}
\big|^2\,
\big|
\braket
  {\LcName (s_{\Lc}^{P_{\Lc}}, |\lambda|)}
  {\LbName (s_{\Lb}^{P_{\Lb}}, |\lambda|)}
\big|^2
\nonumber\\
&\hspace{5mm}
=
\frac{1}{2s_{\Lb}+1}
\sum_{\lambda=-J}^{J} 
\big|
\braket
  {\LcName (s_{\Lc}^{P_{\Lc}}, |\lambda|)}
  {\LbName (s_{\Lb}^{P_{\Lb}}, |\lambda|)}
\big|^2
\,,
\label{eq:CGrelation}
\end{align}
where $J=\mathrm{min}(s_{\Lc},s_{\Lb})$, and the final expression is independent of $J_{H_b}$ and $\lambda_b$.

\section{Decay amplitudes}
\label{sec:decayamplitudes}

In this Appendix, we present the expressions of the decay amplitudes in Eq.~\eqref{eq:decayamplitude2} for each ground-state transition.

For the $B\to D$ transition, the amplitude is written as 
\begin{align}
\mathcal{M}_{D}
&= 
\braket
  {\tfrac{1}{2}, \tfrac{1}{2}\, ; \tfrac{1}{2}, -\tfrac{1}{2}}
  {0, 0}\,
\braket
  {\tfrac{1}{2}, \tfrac{1}{2}\, ; \tfrac{1}{2}, -\tfrac{1}{2}}
  {0, 0}\,
M_{++}\,
\braket
  {\bar{q} (\tfrac{1}{2}^{-}, -\tfrac{1}{2})}
  {\bar{q} (\tfrac{1}{2}^{-}, -\tfrac{1}{2})}
\nonumber\\
&\hspace{5mm}
+
\braket
  {\tfrac{1}{2}, -\tfrac{1}{2}\, ; \tfrac{1}{2}, \tfrac{1}{2}}
  {0, 0}\,
\braket
  {\tfrac{1}{2}, -\tfrac{1}{2}\, ; \tfrac{1}{2}, \tfrac{1}{2}}
  {0, 0}\,
M_{--}\,
\braket
  {\bar{q} (\tfrac{1}{2}^{-}, \tfrac{1}{2})}
  {\bar{q} (\tfrac{1}{2}^{-}, \tfrac{1}{2})}
\,,
\nonumber\\
&=
\frac{1}{2}\,
\bigl(
M_{++}
+
M_{--}
\bigr)
\braket
  {\bar{q} (\tfrac{1}{2}^{-}, \tfrac{1}{2})}
  {\bar{q} (\tfrac{1}{2}^{-}, \tfrac{1}{2})}
\,. 
\end{align}
In the case of $B\to D^*$, the amplitudes for each $D^*$ spin are given by 
\begin{align}
&
\mathcal{M}_{D^*}(\lambda_{D^*})
\nonumber\\
&\hspace{0mm}= 
\left\{
\begin{array}{l}
\braket
  {\tfrac{1}{2}, \tfrac{1}{2}\, ; \tfrac{1}{2}, -\tfrac{1}{2}}
  {1, 0}
\braket
  {\tfrac{1}{2}, \tfrac{1}{2}\, ; \tfrac{1}{2}, -\tfrac{1}{2}}
  {0, 0}\,
M_{++}\,
\braket
  {\bar{q} (\tfrac{1}{2}^{-}, -\tfrac{1}{2})}
  {\bar{q} (\tfrac{1}{2}^{-}, -\tfrac{1}{2})}
\\
\hspace{2mm}
+\,
\braket
  {\tfrac{1}{2}, -\tfrac{1}{2}\, ; \tfrac{1}{2}, \tfrac{1}{2}}
  {1, 0}
\braket
  {\tfrac{1}{2}, -\tfrac{1}{2}\, ; \tfrac{1}{2}, \tfrac{1}{2}}
  {0, 0}\,
M_{--}\,
\braket
  {\bar{q} (\tfrac{1}{2}^{-}, \tfrac{1}{2})}
  {\bar{q} (\tfrac{1}{2}^{-}, \tfrac{1}{2})}
\,,
\\[2mm]
\braket
  {\tfrac{1}{2}, \tfrac{1}{2}\, ; \tfrac{1}{2}, \tfrac{1}{2}}
  {1, 1}
\braket
  {\tfrac{1}{2}, -\tfrac{1}{2}\, ; \tfrac{1}{2}, \tfrac{1}{2}}
  {0, 0}\,
M_{+-}\,
\braket
  {\bar{q} (\tfrac{1}{2}^{-}, \tfrac{1}{2})}
  {\bar{q} (\tfrac{1}{2}^{-}, \tfrac{1}{2})}
\,,
\\[2mm]
\braket
  {\tfrac{1}{2}, -\tfrac{1}{2}\, ; \tfrac{1}{2}, -\tfrac{1}{2}}
  {1, -1}
\braket
  {\tfrac{1}{2}, \tfrac{1}{2}\, ; \tfrac{1}{2}, -\tfrac{1}{2}}
  {0, 0}\,
M_{-+}\,
\braket
  {\bar{q} (\tfrac{1}{2}^{-}, -\tfrac{1}{2})}
  {\bar{q} (\tfrac{1}{2}^{-}, -\tfrac{1}{2})}
\,,
\end{array}
\right.
\nonumber\\[3mm]
&=
\left\{
\begin{array}{l}
\displaystyle
\frac{1}{2}\,
\bigl(
M_{++}
-
M_{--}
\bigr)
\braket
  {\bar{q} (\tfrac{1}{2}^{-}, \tfrac{1}{2})}
  {\bar{q} (\tfrac{1}{2}^{-}, \tfrac{1}{2})}
\,,
\\[3mm]
\displaystyle
-
\frac{1}{\sqrt{2}}\,
M_{+-}\,
\braket
  {\bar{q} (\tfrac{1}{2}^{-}, \tfrac{1}{2})}
  {\bar{q} (\tfrac{1}{2}^{-}, \tfrac{1}{2})}
\,,
\\[3mm]
\displaystyle
\frac{1}{\sqrt{2}}\,
M_{-+}\,
\braket
  {\bar{q} (\tfrac{1}{2}^{-}, \tfrac{1}{2})}
  {\bar{q} (\tfrac{1}{2}^{-}, \tfrac{1}{2})}
\,,
\end{array}
\right.
\end{align}
which correspond to $\lambda_{D^*}=0$, $1$ and $-1$, respectively.
Squaring the amplitudes and summing over $B\to D$ and $B\to D^*$, we obtain 
\begin{align}
\big|
  \mathcal{M}_{D}
\big|^2
+
\sum_{\lambda_{D^*}}
\big|
  \mathcal{M}_{D^*}(\lambda_{D^*})
\big|^2
&=
\frac{1}{2}\,
\Bigl(
\big| M_{++} \big|^2
+
\big| M_{--} \big|^2
+
\big| M_{+-} \big|^2
+
\big| M_{-+} \big|^2
\Bigr)
\nonumber\\
&\hspace{5mm}
\times
\braket
  {\bar{q} (\tfrac{1}{2}^{-}, \tfrac{1}{2})}
  {\bar{q} (\tfrac{1}{2}^{-}, \tfrac{1}{2})}^2
\,, 
\label{eq:ampSquaredBD}
\end{align}
which is consistent with Eq.~\eqref{eq:AmpSquaredGeneral}.

In the $\Lambda_b\to\Lambda_c$ case, the spin of the heavy quark coincides with that of the heavy baryon, yielding
\begin{align}
&
\mathcal{M}_{\Lambda_c}(\lambda_{\Lambda_c},\lambda_{\Lambda_b})
\nonumber\\
&\hspace{5mm}
= 
\braket
  {\tfrac{1}{2}, \lambda_{\Lambda_c}\, ; 0, 0}
  {\tfrac{1}{2}, \lambda_{\Lambda_c}}
\braket
  {\tfrac{1}{2}, \lambda_{\Lambda_b}\, ; 0, 0}
  {\tfrac{1}{2}, \lambda_{\Lambda_b}}\,
M_{\lambda_{\Lambda_c},\lambda_{\Lambda_b}}\,
\braket
  {qq (0^{+}, 0)}
  {qq (0^{+}, 0)}
\,,
\nonumber\\
&\hspace{5mm}
= 
M_{\lambda_{\Lambda_c},\lambda_{\Lambda_b}}
\braket
  {qq (0^{+},0)}
  {qq (0^{+},0)}
\,.
\end{align}
The squared amplitude is given by 
\begin{align}
\frac12
\sum_{\lambda_{\Lambda_{c}},\lambda_{\Lambda_{b}}}
\big|
  \mathcal{M}_{\Lambda_c}(\lambda_{\Lambda_{c}},\lambda_{\Lambda_{b}})
\big|^2
&=
\frac{1}{2}\,
\Bigl(
\big| M_{++} \big|^2
+
\big| M_{--} \big|^2
+
\big| M_{+-} \big|^2
+
\big| M_{-+} \big|^2
\Bigr)
\nonumber\\
&\hspace{5mm}
\times
\braket
  {qq (0^{+},0)}
  {qq (0^{+},0)}^2
\,,
\label{eq:ampSquaredLambda}
\end{align}
where the factor $1/2$ reflects the average over the spin of the initial baryon.

The amplitudes for the $\Omega_b\to\Omega_c$ transition are given by 
\begin{align}
&
\mathcal{M}_{\Omega_c}(\lambda_{\Omega_c},\lambda_{\Omega_b})
\nonumber\\
&\hspace{0mm}= 
\left\{
\begin{array}{ll}
\braket
  {\tfrac{1}{2}, \tfrac{1}{2}\, ; 1, 0}
  {\tfrac{1}{2}, \tfrac{1}{2}}
\braket
  {\tfrac{1}{2}, \tfrac{1}{2}\, ; 1, 0}
  {\tfrac{1}{2}, \tfrac{1}{2}}\,
M_{++}\,
\braket
  {ss(1^{+},0)}
  {ss(1^{+},0)}
\\
\hspace{2mm}
+\,
\braket
  {\tfrac{1}{2}, -\tfrac{1}{2}\, ; 1, 1}
  {\tfrac{1}{2}, \tfrac{1}{2}}
\braket
  {\tfrac{1}{2}, -\tfrac{1}{2}\, ; 1, 1}
  {\tfrac{1}{2}, \tfrac{1}{2}}\,
M_{--}\,
\braket
  {ss(1^{+},1)}
  {ss(1^{+},1)}
\,,
\\[2mm]
\braket
  {\tfrac{1}{2}, -\tfrac{1}{2}\, ; 1, 0}
  {\tfrac{1}{2}, -\tfrac{1}{2}}
\braket
  {\tfrac{1}{2}, -\tfrac{1}{2}\, ; 1, 0}
  {\tfrac{1}{2}, -\tfrac{1}{2}}\,
M_{--}\,
\braket
  {ss(1^{+},0)}
  {ss(1^{+},0)}
\\
\hspace{2mm}
+\,
\braket
  {\tfrac{1}{2}, \tfrac{1}{2}\, ; 1, -1}
  {\tfrac{1}{2}, -\tfrac{1}{2}}
\braket
  {\tfrac{1}{2}, \tfrac{1}{2}\, ; 1, -1}
  {\tfrac{1}{2}, -\tfrac{1}{2}}\,
M_{++}\,
\braket
  {ss(1^{+},-1)}
  {ss(1^{+},-1)}
\,,
\\[2mm]
\braket
  {\tfrac{1}{2}, \tfrac{1}{2}\, ; 1, 0}
  {\tfrac{1}{2}, \tfrac{1}{2}}
\braket
  {\tfrac{1}{2}, -\tfrac{1}{2}\, ; 1, 0}
  {\tfrac{1}{2}, -\tfrac{1}{2}}\,
M_{+-}\,
\braket
  {ss(1^{+},0)}
  {ss(1^{+},0)}
\,,
\\[2mm]
\braket
  {\tfrac{1}{2}, -\tfrac{1}{2}\, ; 1, 0}
  {\tfrac{1}{2}, -\tfrac{1}{2}}
\braket
  {\tfrac{1}{2}, \tfrac{1}{2}\, ; 1, 0}
  {\tfrac{1}{2}, \tfrac{1}{2}}\,
M_{-+}\,
\braket
  {ss(1^{+},0)}
  {ss(1^{+},0)}
\,,
\nonumber
\end{array}
\right.
\nonumber\\[3mm]
&=
\left\{
\begin{array}{l}
\displaystyle
\frac{1}{3}\,
\Big(
M_{++}\,
\braket
  {ss(1^{+},0)}
  {ss(1^{+},0)}
+
2\,
M_{--}\,
\braket
  {ss(1^{+},1)}
  {ss(1^{+},1)}
\Big)
\,,
\\[3mm]
\displaystyle
\frac{1}{3}\,
\Big(
M_{--}\,
\braket
  {ss(1^{+},0)}
  {ss(1^{+},0)}
+
2\,
M_{++}\,
\braket
  {ss(1^{+},1)}
  {ss(1^{+},1)}
\Big)
\,,
\\[3mm]
\displaystyle
-\frac{1}{3}\,
M_{+-}\,
\braket
  {ss(1^{+},0)}
  {ss(1^{+},0)}
\,,
\\[3mm]
\displaystyle
-\frac{1}{3}\,
M_{-+}\,
\braket
  {ss(1^{+},0)}
  {ss(1^{+},0)}
\,,
\end{array}
\right.
\end{align}
for 
$(\lambda_{\Omega_c},\lambda_{\Omega_b})=(\tfrac12,\tfrac12)$, $(-\tfrac12,-\tfrac12)$, $(\tfrac12,-\tfrac12)$ and $(-\tfrac12,\tfrac12)$, respectively, while those for the $\Omega_b\to\Omega^*_c$ transition are given by 
\begin{align}
&
\mathcal{M}_{\Omega^*_c}(\lambda_{\Omega^*_c},\lambda_{\Omega_b})
\nonumber\\
&\hspace{0mm}= 
\left\{
\begin{array}{l}
\braket
  {\tfrac{1}{2}, \tfrac{1}{2}\, ; 1, 0}
  {\tfrac{3}{2}, \tfrac{1}{2}}
\braket
  {\tfrac{1}{2}, \tfrac{1}{2}\, ; 1, 0}
  {\tfrac{1}{2}, \tfrac{1}{2}}\,
M_{++}\,
\braket
  {ss(1^{+},0)}
  {ss(1^{+},0)}
\\
\hspace{2mm}
+\,
\braket
  {\tfrac{1}{2}, -\tfrac{1}{2}\, ; 1, 1}
  {\tfrac{3}{2}, \tfrac{1}{2}}
\braket
  {\tfrac{1}{2}, -\tfrac{1}{2}\, ; 1, 1}
  {\tfrac{1}{2}, \tfrac{1}{2}}\,
M_{--}\,
\braket
  {ss(1^{+},1)}
  {ss(1^{+},1)}
\,,
\\[2mm]
\braket
  {\tfrac{1}{2}, -\tfrac{1}{2}\, ; 1, 0}
  {\tfrac{3}{2}, -\tfrac{1}{2}}
\braket
  {\tfrac{1}{2}, -\tfrac{1}{2}\, ; 1, 0}
  {\tfrac{1}{2}, -\tfrac{1}{2}}\,
M_{--}\,
\braket
  {ss(1^{+},0)}
  {ss(1^{+},0)}
\\
\hspace{2mm}
+\,
\braket
  {\tfrac{1}{2}, \tfrac{1}{2}\, ; 1, -1}
  {\tfrac{3}{2}, -\tfrac{1}{2}}
\braket
  {\tfrac{1}{2}, \tfrac{1}{2}\, ; 1, -1}
  {\tfrac{1}{2}, -\tfrac{1}{2}}\,
M_{++}\,
\braket
  {ss(1^{+},-1)}
  {ss(1^{+},-1)}
\,,
\\[2mm]
\braket
  {\tfrac{1}{2}, \tfrac{1}{2}\, ; 1, 0}
  {\tfrac{3}{2}, \tfrac{1}{2}}
\braket
  {\tfrac{1}{2}, -\tfrac{1}{2}\, ; 1, 0}
  {\tfrac{1}{2}, -\tfrac{1}{2}}\,
M_{+-}\,
\braket
  {ss(1^{+},0)}
  {ss(1^{+},0)}
\,,
\\[2mm]
\braket
  {\tfrac{1}{2}, -\tfrac{1}{2}\, ; 1, 0}
  {\tfrac{3}{2}, -\tfrac{1}{2}}
\braket
  {\tfrac{1}{2}, \tfrac{1}{2}\, ; 1, 0}
  {\tfrac{1}{2}, \tfrac{1}{2}}\,
M_{-+}\,
\braket
  {ss(1^{+},0)}
  {ss(1^{+},0)}
\,,
\\[2mm]
\braket
  {\tfrac{1}{2}, \tfrac{1}{2}\, ; 1, 1}
  {\tfrac{3}{2}, \tfrac{3}{2}}
\braket
  {\tfrac{1}{2}, -\tfrac{1}{2}\, ; 1, 1}
  {\tfrac{1}{2}, \tfrac{1}{2}}\,
M_{+-}\,
\braket
  {ss(1^{+},1)}
  {ss(1^{+},1)}
\,,
\\[2mm]
\braket
  {\tfrac{1}{2}, -\tfrac{1}{2}\, ; 1, -1}
  {\tfrac{3}{2}, -\tfrac{3}{2}}
\braket
  {\tfrac{1}{2}, \tfrac{1}{2}\, ; 1, -1}
  {\tfrac{1}{2}, -\tfrac{1}{2}}\,
M_{-+}\,
\braket
  {ss(1^{+},-1)}
  {ss(1^{+},-1)}
\,,
\end{array}
\right.
\nonumber\\[3mm]
&=
\left\{
\begin{array}{l}
\displaystyle
\frac{\sqrt{2}}{3}\,
\Big(
M_{++}\,
\braket
  {ss(1^{+},0)}
  {ss(1^{+},0)}
-
M_{--}\,
\braket
  {ss(1^{+},1)}
  {ss(1^{+},1)}
\Big)
\,,
\\[3mm]
\displaystyle
-\frac{\sqrt{2}}{3}\,
\Big(
M_{--}\,
\braket
  {ss(1^{+},0)}
  {ss(1^{+},0)}
-
M_{++}\,
\braket
  {ss(1^{+},1)}
  {ss(1^{+},1)}
\Big)
\,,
\\[3mm]
\displaystyle
-\frac{\sqrt{2}}{3}\,
M_{+-}\,
\braket
  {ss(1^{+},0)}
  {ss(1^{+},0)}
\,,
\\[3mm]
\displaystyle
\frac{\sqrt{2}}{3}\,
M_{-+}\,
\braket
  {ss(1^{+},0)}
  {ss(1^{+},0)}
\,,
\\[3mm]
\displaystyle
-\sqrt{\frac{2}{3}}\,
M_{+-}\,
\braket
  {ss(1^{+},1)}
  {ss(1^{+},1)}
\,,
\\[3mm]
\displaystyle
\sqrt{\frac{2}{3}}\,
M_{-+}\,
\braket
  {ss(1^{+},1)}
  {ss(1^{+},1)}
\,,
\end{array}
\right.
\end{align}
for 
$(\lambda_{\Omega_c^*},\lambda_{\Omega_b})=(\tfrac12,\tfrac12)$, $(-\tfrac12,-\tfrac12)$, $(\tfrac12,-\tfrac12)$, $(-\tfrac12,\tfrac12)$, $(\tfrac32,\tfrac12)$ and $(-\tfrac32,-\tfrac12)$, respectively,
Averaging over the spin of the initial-state baryon and summing over that of the final-state baryon, the sum of the squared amplitudes becomes 
\begin{align}
&
\frac12
\sum_{\lambda_{\Omega_{c}},\lambda_{\Omega_{b}}}
\big|
  \mathcal{M}_{\Omega_{c}}(\lambda_{\Omega_{c}},\lambda_{\Omega_{b}})
\big|^2
+
\frac12
\sum_{\lambda_{\Omega_{c}^*},\lambda_{\Omega_{b}}}
\big|
  \mathcal{M}_{\Omega_{c}^*}(\lambda_{\Omega_{c}^*},\lambda_{\Omega_{b}})
\big|^2
\nonumber\\
&\hspace{5mm}
=
\frac{1}{2}\,
\Bigl(
\big| M_{++} \big|^2
+
\big| M_{--} \big|^2
+
\big| M_{+-} \big|^2
+
\big| M_{-+} \big|^2
\Bigr)
\nonumber\\[1mm]
&\hspace{10mm}
\times
\frac13\,
\Big(
\braket
  {ss(1^{+},0)}
  {ss(1^{+},0)}^2
+
2\,
\braket
  {ss(1^{+},1)}
  {ss(1^{+},1)}^2
\Big)
\,, 
\label{eq:ampSquaredOmega}
\end{align}
which has the same structure as Eqs.~\eqref{eq:ampSquaredBD} and \eqref{eq:ampSquaredLambda} apart from the matrix element of the light component.

\section{Isgur--Wise functions}
\label{sec:IWfunctions}

For completeness, we summarize the definitions of the IW functions and their relation to the light-component matrix elements in the spin decomposition picture. 
This clarifies the equivalence between the usual form-factor parametrization and the spin decomposition picture.

The $B\to D$ transition matrix element of the vector current is expressed 
in terms of the IW function $\xi(w)$ as\cite{Isgur:1989vq,Isgur:1990yhj} 
\begin{align}
\braket
  {D}
  {\bar{c}\,\gamma^{\mu} b}
  {B}
&=
\sqrt{m_B m_D}\, 
\xi(w)\,(v+v')^\mu
\,,
\label{eq:BDvectorIW}
\end{align}
where $m_B$ and $m_D$ denote the masses of the $B$ and $D$ mesons, 
$v$ and $v'$ denote their four-velocities, respectively, and $w=v\cdot v'$. 
The normalization of $\xi(w)$ is given by $\xi(1)=1$. 
Using Eq.~\eqref{eq:hadronic}, the same matrix element can be rewritten in the spin decomposition picture with the Clebsch--Gordan coefficients, the heavy-quark spinors and the overlap of the light degrees of freedom as
\begin{align}
\braket
  {D}
  {\bar{c}\,\gamma^{\mu} b}
  {B}
&=
\braket
  {\tfrac12, \tfrac12\, ; \tfrac12, -\tfrac12}
  {0,0}^2\,
\bar{u}_c(\tfrac12) \gamma^\mu u_b(\tfrac12)\,
\braket
  {\bar{q} (\tfrac12^-, -\tfrac12)}
  {\bar{q} (\tfrac12^-, -\tfrac12)}
\nonumber\\
&\hspace{5mm}
+
\braket
  {\tfrac12, -\tfrac12\, ; \tfrac12, \tfrac12}
  {0,0}^2\,
\bar{u}_c(-\tfrac12) \gamma^\mu u_b(-\tfrac12)\,
\braket
  {\bar{q} (\tfrac12^-, \tfrac12)}
  {\bar{q} (\tfrac12^-, \tfrac12)}
\,,
\nonumber\\
&=
\frac12\,
\Big[
\bar{u}_c(\tfrac12) \gamma^\mu u_b(\tfrac12)
+
\bar{u}_c(-\tfrac12) \gamma^\mu u_b(-\tfrac12)
\Big]
\braket
  {\bar{q} (\tfrac12^-, \tfrac12)}
  {\bar{q} (\tfrac12^-, \tfrac12)}
\,,
\nonumber\\
&=
\sqrt{\frac{2\, m_B m_D}{w+1}}\,
\big( v + v' \big)^\mu
\braket
  {\bar{q} (\tfrac12^-, \tfrac12)}
  {\bar{q} (\tfrac12^-, \tfrac12)}
\,. 
\label{eq:BDvector}
\end{align}
In the last equality, we used $\bar{u}_c(\pm\tfrac12) \gamma^\mu u_b(\pm\tfrac12)=\!\sqrt{2m_B m_D/(w+1)}\,(v+v')^\mu$. 
Comparing Eqs.~\eqref{eq:BDvector} and \eqref{eq:BDvectorIW}, we obtain the relation,
\begin{align}
\xi(w)
&=
\sqrt{\frac{2}{w+1}}\,
\braket
  {\bar{q} (\tfrac{1}{2}^{-}, \tfrac{1}{2})}
  {\bar{q} (\tfrac{1}{2}^{-}, \tfrac{1}{2})}
\,.
\end{align}
This is given in Eq.~\eqref{eq:xi}. 
The same IW function is obtained by the other currents related to the vector one under the heavy-quark symmetry.
This result is also obtained correspondingly in the spin decomposition picture. 
The same argument also holds for the $B\to D^*$ transition.

For the $\Lambda_b\to\Lambda_c$ transition, the IW function $\zeta(w)$ is defined as~\cite{Isgur:1990pm,Georgi:1990cx} 
\begin{align}
\braket
  {\Lambda_c(\lambda_{\Lambda_c})}
  {\bar{c}\,\Gamma\,b}
  {\Lambda_b(\lambda_{\Lambda_b})}
= 
\zeta(w)\, 
\bar{u}_{\Lambda_c}(\lambda_{\Lambda_c})\, \Gamma\, u_{\Lambda_b}(\lambda_{\Lambda_b}) 
\,,
\end{align}
where $\Gamma$ denotes a general $\gamma$ matrix, and $\zeta(1)=1$. 
In the spin decomposition picture, this matrix element takes the form 
\begin{align}
\braket
  {\Lambda_c(\lambda_{\Lambda_c})}
  {\bar{c}\,\Gamma\,b}
  {\Lambda_b(\lambda_{\Lambda_b})}
&=
\braket
  {\tfrac{1}{2},\lambda_{\Lambda_c}\, ; 0, 0}
  {\tfrac{1}{2}, \lambda_{\Lambda_c}}
\braket
  {\tfrac{1}{2},\lambda_{\Lambda_b}\, ; 0, 0}
  {\tfrac{1}{2}, \lambda_{\Lambda_b}}
\nonumber\\
&\hspace{10mm}\times
\bar{u}_c(\lambda_{\Lambda_c})\,\Gamma\, u_b(\lambda_{\Lambda_b})\,
\braket
  {qq (0^+,0)}
  {qq (0^+,0)}
\,
\nonumber\\
&=
\bar{u}_c(\lambda_{\Lambda_c})\,\Gamma\, u_b(\lambda_{\Lambda_b})\,
\braket
  {qq (0^+,0)}
  {qq (0^+,0)}
\,.
\end{align}
Since the spinors of the heavy quarks coincide with those of the baryons in the heavy-quark limit, we obtain the relation,
\begin{align}
\zeta(w)
&=
\braket
  {qq (0^{+},0)}
  {qq (0^{+},0)}
\,. 
\end{align}
This is given in Eq.~\eqref{eq:zeta}. 

On the other hand, the $\Omega_b\to\Omega_c$ transition is described by two IW functions $\xi_1(w)$ and $\xi_2(w)$, defined as~\cite{Isgur:1990pm,Georgi:1990cx}  
\begin{align}
\braket
  {\Omega_c(\lambda_{\Omega_c})}
  {\bar{c}\,\Gamma\,b}
  {\Omega_b(\lambda_{\Omega_b})}
=
\bigl[
  -g_{\mu\nu}\, \xi_1(w) + v_\mu v'_\nu\, \xi_2(w)
\bigr]\,
\overline{\Omega}_c^{\,\mu}(\lambda_{\Omega_c})\, \Gamma\, 
\Omega_b^{\,\nu}(\lambda_{\Omega_b})
\,,
\end{align}
where $\xi_1(1)=1$, and $\Omega_b^{\,\nu}(\lambda_{\Omega_b})$ is given in terms of the Dirac spinor $u_{\Omega_b}(\lambda_{\Omega_b})$ as 
\begin{align}
\Omega_b^{\,\nu}(\lambda_{\Omega_b}) 
&= 
\frac{1}{\sqrt{3}} 
\bigl( \gamma^\nu + v^\nu \bigr) \gamma_5\, 
u_{\Omega_b}(\lambda_{\Omega_b})
\,.
\end{align}
For example, in the case of $\Gamma=\gamma^\rho$, we obtain 
\begin{align}
\braket
  {\Omega_c(\pm\tfrac12)}
  {\bar{c}\,\gamma^\rho b}
  {\Omega_b(\pm\tfrac12)}
&=
\frac{1}{3}\,
\bigl[
(w+2)\, \xi_1(w)
-
(w^2-1)\, \xi_2(w)
\bigr]\,
\bar{u}_{\Omega_c}(\pm\tfrac12) \gamma^\rho u_{\Omega_b}(\pm\tfrac12)
\,,
\label{eq:IWfunctionsOmega1}
\\
\braket
  {\Omega_c(\pm\tfrac12)}
  {\bar{c}\,\gamma^\rho b}
  {\Omega_b(\mp\tfrac12)}
&=
\frac{1}{3}\,
\bigl[
-
w\, \xi_1(w)
+
(w^2-1)\, \xi_2(w)
\bigr]\,
\bar{u}_{\Omega_c}(\pm\tfrac12) \gamma^\rho u_{\Omega_b}(\mp\tfrac12)
\,,
\label{eq:IWfunctionsOmega2}
\end{align}
The corresponding matrix elements in the spin decomposition picture are given by 
\begin{align}
&
\braket
  {\Omega_c(\pm\tfrac12)}
  {\bar{c}\,\gamma^{\rho} b}
  {\Omega_b(\pm\tfrac12)}
\nonumber\\
&\hspace{5mm}
=
\braket
  {\tfrac{1}{2}, \pm\tfrac{1}{2}\, ; 1, 0}
  {\tfrac{1}{2}, \pm\tfrac{1}{2}}^2\,
\bar{u}_{c}(\pm\tfrac{1}{2})\gamma^\rho u_b(\pm\tfrac{1}{2})\,
\braket
  {ss(1^+,0)}
  {ss(1^+,0)}
\nonumber\\
&\hspace{10mm}
+
\braket
  {\tfrac{1}{2}, \mp\tfrac{1}{2}\, ; 1, \pm 1}
  {\tfrac{1}{2}, \pm\tfrac{1}{2}}^2\,
\bar{u}_{c}(\mp\tfrac{1}{2})\gamma^\rho u_b(\mp\tfrac{1}{2})\,
\braket
  {ss(1^+,\pm 1)}
  {ss(1^+,\pm 1)}
\nonumber\\
&\hspace{5mm}
=
\frac{1}{3}\,
\bar{u}_{c}(\pm\tfrac{1}{2})\gamma^\rho u_b(\pm\tfrac{1}{2})
\Bigl[
\braket
  {ss(1^+,0)}
  {ss(1^+,0)}
+
2\,
\braket
  {ss(1^+,\pm 1)}
  {ss(1^+,\pm 1)}
\Bigr]
\,,\!
\label{eq:IWfunctionsOmega3}
\\
&
\braket
  {\Omega_c(\pm\tfrac12)}
  {\bar{c}\,\gamma^{\rho} b}
  {\Omega_b(\mp\tfrac12)}
\nonumber\\
&\hspace{5mm}
=
\braket
  {\tfrac{1}{2}, \pm\tfrac{1}{2}\, ; 1, 0}
  {\tfrac{1}{2}, \pm\tfrac{1}{2}}
\braket
  {\tfrac{1}{2}, \mp\tfrac{1}{2}\, ; 1, 0}
  {\tfrac{1}{2}, \mp\tfrac{1}{2}}\,
\bar{u}_{c}(\pm\tfrac{1}{2})\gamma^\rho u_b(\mp\tfrac{1}{2})\,
\braket
  {ss(1^+,0)}
  {ss(1^+,0)}
\,,
\nonumber\\
&\hspace{5mm}
=
-\frac{1}{3}\,
\bar{u}_{c}(\pm\tfrac{1}{2})\gamma^\rho u_b(\mp\tfrac{1}{2})\,
\braket
  {ss(1^+,0)}
  {ss(1^+,0)}
\,.
\label{eq:IWfunctionsOmega4}
\end{align}
From Eqs.~\eqref{eq:IWfunctionsOmega1}-\eqref{eq:IWfunctionsOmega4}, we find that the IW functions are written in terms of the matrix elements of the light degrees of freedom as
\begin{align}
\xi_1(w) 
&=
\braket
  {ss(1^{+},1)}
  {ss(1^{+},1)}
\,,
\\
\xi_2(w) 
&=
\frac{1}{w^2-1}
\Big(
w\,
\braket
  {ss(1^{+},1)}
  {ss(1^{+},1)}
-
\braket
  {ss(1^{+},0)}
  {ss(1^{+},0)}
\Big)
\,.
\end{align}
They are the same as Eqs.~\eqref{eq:IW_xi1} and \eqref{eq:IW_xi2}. 

Finally, the IW functions $\tau_{1/2}(w)$ and $\tau_{3/2}(w)$ for the $B\to \{D_0^*,D_1^*\}$ and $B\to \{D_1,D_2^*\}$ transitions are defined, for instance, through the vector-current matrix elements as~\cite{Isgur:1990jf} 
\begin{align}
&
  \braket
  {D^{*}_{1}(\lambda_{D^{*}_{1}})}
  {\bar{c}\,\gamma^{\mu} b}
  {B}
=
2
\sqrt{ \smash[b]{ m_{B} m_{D^{*}_{1}} } }\,
\bigl[
(w-1)\, \epsilon^{\ast\mu}
-  
(\epsilon^{*}\cdot v)\,
v^{\prime\mu}
\bigr]\,
\tau_{1/2}(w)
\,,
\\  
&
\braket
  {D_{1}(\lambda_{D_{1}})}
  {\bar{c}\,\gamma^{\mu} b}
  {B}
\nonumber\\
&\hspace{10mm}=
\sqrt{ \frac{m_{B} m_{D_{1}}}{2} }\,
\bigl[
(1-w^2)\,
\epsilon^{\ast\mu} 
+
(\epsilon^{*}\cdot v)
\bigl(
-
3\, v^\mu
+
(w-2)\, v^{\prime\mu}
\bigr)
\bigr]\,
\tau_{3/2}(w)
\,,
\end{align}
where $m_{D^{*}_{1}}$ and $m_{D_{1}}$ are the masses of the $D^{*}_{1}$ and $D_{1}$ mesons, respectively, and $\epsilon^{\ast\mu}$ denotes the polarization vector of the charm mesons. 
As in the previous cases, by comparing these expressions with those obtained in the spin decomposition picture, we obtain the relations given in Eqs.~\eqref{eq:IW_tau12} and \eqref{eq:IW_tau32} as
\begin{align}
\tau_{1/2}(w)
&=
\frac{1}{\sqrt{2(w-1)}}\,
\braket
  {\bar{q} (\tfrac{1}{2}^{+}, \tfrac{1}{2})}
  {\bar{q} (\tfrac{1}{2}^{-}, \tfrac{1}{2})}
\,,
\\
\tau_{3/2}(w)
&=
-
\frac{1}{(w+1)}
\frac{1}{\sqrt{w-1}}\,
\braket
  {\bar{q} (\tfrac{3}{2}^{+}, \tfrac{1}{2})}
  {\bar{q} (\tfrac{1}{2}^{-}, \tfrac{1}{2})}
\,. 
\end{align}
The same relations also hold for other matrix elements in the $B\to \{D_0^*,D_1^*\}$ and $B\to \{D_1,D_2^*\}$ transitions.

For each transition, the hadronic matrix elements can be expressed either in terms of the IW functions or as overlaps of the light components in the spin decomposition picture, leading to identical results. 
We thus conclude that the spin decomposition picture provides an equivalent description of the hadronic matrix elements as that given by the IW functions.

\section{Bjorken sum rules}
\label{sec:BjorkenSumRules}

Here, we briefly comment on the derivation of the Bjorken sum rules~\cite{Bjorken:1990hs,Isgur:1990jf} in terms of the hadron decay amplitude obtained in Eq.~\eqref{eq:AmpSquaredGeneral}. 

For the $B$-meson semileptonic decays $B\to H_c\hspace{0.08em}l\hspace{0.08em}\bar{\nu}_l$, summing the squared decay amplitudes in the spin decomposition picture over all possible final-state mesons $H_c$ and integrating over the phase space, we obtain the inclusive differential decay rate 
\begin{align}
  &
  \frac{d\hspace{0.08em}\Gamma(B\to X_c\hspace{0.08em}l\hspace{0.08em}\bar{\nu}_l)}{dw}
  =
  \sum_{H_c}
  \frac{d\hspace{0.08em}\Gamma(B\to H_c\hspace{0.08em}l\hspace{0.08em}\bar{\nu}_l)}{dw}
  \label{eq:inclusive}
  \\
  &\hspace{5mm}=
  \frac{d\hspace{0.08em}\Gamma(b\to c\hspace{0.08em}l\hspace{0.08em}\bar{\nu}_l)}{dw}
  \bigg[
    \frac{1+w}{2}\,\xi(w)^2
    +
    (w - 1)
    \Big(
      2\,\tau_{1/2}(w)^2
      +
      (w + 1)^2\,\tau_{3/2}(w)^2
    \Big)
    + \cdots
  \bigg]\,,
  \nonumber
\end{align}
where $d\hspace{0.08em}\Gamma(b\to c\hspace{0.08em}l\hspace{0.08em}\bar{\nu}_l)/dw$ is the partonic differential rate, and the ellipsis denotes contributions from higher excited states. 
Here, we applied Eq.~\eqref{eq:AmpSquaredGeneral} in the second equality. 
By assuming the quark-hadron duality, Eq.~\eqref{eq:inclusive} yields the Bjorken sum rule~\cite{Bjorken:1990hs,Isgur:1990jf} 
\begin{align}
  1
  =
  \frac{1+w}{2}\,\xi(w)^2
  +
  (w - 1)
  \Big(
    2\,\tau_{1/2}(w)^2
    +
    (w + 1)^2\,\tau_{3/2}(w)^2
  \Big)
  + \cdots
  \,. 
\end{align}
Analogous Bjorken sum rules also hold for the semileptonic $\Lambda_b$ decays~\cite{Isgur:1991wr} and the semileptonic $\Omega_b$ decays~\cite{Xu:1993mj,Hussain:1993qd}.

\bibliographystyle{utphys28mod}
\bibliography{ref}

\end{document}